
\input phyzzx.tex
%
\catcode`\@=11
\def\wlog#1{}
\def\eqname#1{\rel@x {\pr@tect
  \ifnum\equanumber<0 \xdef#1{{\rm\number-\equanumber}}%
     \gl@bal\advance\equanumber by -1
  \else \gl@bal\advance\equanumber by 1
     \ifx\chapterlabel\rel@x \def\d@t{}\else \def\d@t{.}\fi
    \xdef#1{{\rm\chapterlabel\d@t\number\equanumber}}\fi #1}}
\catcode`\@=12
\catcode`\@=11

\def\eat@#1{}
\mathchardef\prime@="0230
\def\prime{{{}\prime@{}}}
\def\prim@s{\prime@\futurelet\next\pr@m@s}

\def\,{\relax\ifmmode\mskip\thinmuskip\else\thinspace\fi}
\def\!{\relax\ifmmode\mskip-\thinmuskip\else\negthinspace\fi}
\def\frac#1#2{{#1\over#2}}

\def\:{\nobreak\hskip.1111em{:}\hskip.3333em plus .0555em\relax}
\def\intic@{\mathchoice{\hskip5\p@}{\hskip4\p@}{\hskip4\p@}{\hskip4\p@}}
\def\negintic@
 {\mathchoice{\hskip-5\p@}{\hskip-4\p@}{\hskip-4\p@}{\hskip-4\p@}}
\def\intkern@{\mathchoice{\!\!\!}{\!\!}{\!\!}{\!\!}}
\def\intdots@{\mathchoice{\cdots}{{\cdotp}\mkern1.5mu
    {\cdotp}\mkern1.5mu{\cdotp}}{{\cdotp}\mkern1mu{\cdotp}\mkern1mu
      {\cdotp}}{{\cdotp}\mkern1mu{\cdotp}\mkern1mu{\cdotp}}}
\newcount\intno@
\def\iint{\intno@=\tw@\futurelet\next\ints@}
\def\iiint{\intno@=\thr@@\futurelet\next\ints@}
\def\iiiint{\intno@=4 \futurelet\next\ints@}
\def\idotsint{\intno@=\z@\futurelet\next\ints@}
\def\ints@{\findlimits@\ints@@}
\newif\iflimtoken@
\newif\iflimits@
\def\findlimits@{\limtoken@false\limits@false\ifx\next\limits
 \limtoken@true\limits@true\else\ifx\next\nolimits\limtoken@true\limits@false
    \fi\fi}
\def\multintlimits@{\intop\ifnum\intno@=\z@\intdots@
  \else\intkern@\fi
    \ifnum\intno@>\tw@\intop\intkern@\fi
     \ifnum\intno@>\thr@@\intop\intkern@\fi\intop}
\def\multint@{\int\ifnum\intno@=\z@\intdots@\else\intkern@\fi
   \ifnum\intno@>\tw@\int\intkern@\fi
    \ifnum\intno@>\thr@@\int\intkern@\fi\int}
\def\ints@@{\iflimtoken@\def\ints@@@{\iflimits@
   \negintic@\mathop{\intic@\multintlimits@}\limits\else
    \multint@\nolimits\fi\eat@}\else
     \def\ints@@@{\multint@\nolimits}\fi\ints@@@}
\def\Sb{_\bgroup\vspace@
        \baselineskip=\fontdimen10 \scriptfont\tw@
        \advance\baselineskip by \fontdimen12 \scriptfont\tw@
        \lineskip=\thr@@\fontdimen8 \scriptfont\thr@@
        \lineskiplimit=\thr@@\fontdimen8 \scriptfont\thr@@
        \Let@\vbox\bgroup\halign\bgroup \hfil$\scriptstyle
            {##}$\hfil\cr}
\def\endSb{\crcr\egroup\egroup\egroup}
\def\Sp{^\bgroup\vspace@
        \baselineskip=\fontdimen10 \scriptfont\tw@
        \advance\baselineskip by \fontdimen12 \scriptfont\tw@
        \lineskip=\thr@@\fontdimen8 \scriptfont\thr@@
        \lineskiplimit=\thr@@\fontdimen8 \scriptfont\thr@@
        \Let@\vbox\bgroup\halign\bgroup \hfil$\scriptstyle
            {##}$\hfil\cr}
\def\endSp{\crcr\egroup\egroup\egroup}
\def\Let@{\relax\iffalse{\fi\let\\=\cr\iffalse}\fi}
\def\vspace@{\def\vspace##1{\noalign{\vskip##1 }}}
\def\aligned{\,\vcenter\bgroup\vspace@\Let@\openup\jot\m@th\ialign
  \bgroup \strut\hfil$\displaystyle{##}$&$\displaystyle{{}##}$\hfil\crcr}
\def\endaligned{\crcr\egroup\egroup}
\def\matrix{\,\vcenter\bgroup\Let@\vspace@
    \normalbaselines
  \m@th\ialign\bgroup\hfil$##$\hfil&&\quad\hfil$##$\hfil\crcr
    \mathstrut\crcr\noalign{\kern-\baselineskip}}
\def\endmatrix{\crcr\mathstrut\crcr\noalign{\kern-\baselineskip}\egroup
                \egroup\,}
\newtoks\hashtoks@
\hashtoks@={#}
\def\format{\crcr\egroup\iffalse{\fi\ifnum`}=0 \fi\format@}
\def\format@#1\\{\def\preamble@{#1}%
  \def\c{\hfil$\the\hashtoks@$\hfil}%
  \def\r{\hfil$\the\hashtoks@$}%
  \def\l{$\the\hashtoks@$\hfil}%
  \setbox\z@=\hbox{\xdef\Preamble@{\preamble@}}\ifnum`{=0 \fi\iffalse}\fi
   \ialign\bgroup\span\Preamble@\crcr}

\def\cases{\left\{\,\vcenter\bgroup\vspace@
     \normalbaselines\openup\jot\m@th
       \Let@\ialign\bgroup$##$\hfil&\quad$##$\hfil\crcr
      \mathstrut\crcr\noalign{\kern-\baselineskip}}

\newif\iftagsleft@
\tagsleft@true
\def\TagsOnRight{\global\tagsleft@false}
\def\tag#1$${\iftagsleft@\leqno\else\eqno\fi
 \hbox{\def\pagebreak{\global\postdisplaypenalty-\@M}%
 \def\nopagebreak{\global\postdisplaypenalty\@M}\rm(#1\unskip)}%
  $$\postdisplaypenalty\z@\ignorespaces}
\interdisplaylinepenalty=\@M
\def\allowdisplaybreak@{\def\allowdisplaybreak{\noalign{\allowbreak}}}
\def\displaybreak@{\def\displaybreak{\noalign{\break}}}
\def\align#1\endalign{\def\tag{&}\vspace@\allowdisplaybreak@\displaybreak@
  \iftagsleft@\lalign@#1\endalign\else
   \ralign@#1\endalign\fi}
\def\ralign@#1\endalign{\displ@y\Let@\tabskip\centering\halign to\displaywidth
     {\hfil$\displaystyle{##}$\tabskip=\z@&$\displaystyle{{}##}$\hfil
       \tabskip=\centering&\llap{\hbox{(\rm##\unskip)}}\tabskip\z@\crcr
             #1\crcr}}
\def\lalign@
 #1\endalign{\displ@y\Let@\tabskip\centering\halign to \displaywidth
   {\hfil$\displaystyle{##}$\tabskip=\z@&$\displaystyle{{}##}$\hfil
   \tabskip=\centering&\kern-\displaywidth
        \rlap{\hbox{(\rm##\unskip)}}\tabskip=\displaywidth\crcr
               #1\crcr}}
\def\overrightarrow{\mathpalette\overrightarrow@}
\def\overrightarrow@#1#2{\vbox{\ialign{$##$\cr
    #1{-}\mkern-6mu\cleaders\hbox{$#1\mkern-2mu{-}\mkern-2mu$}\hfill
     \mkern-6mu{\to}\cr
     \noalign{\kern -1\p@\nointerlineskip}
     \hfil#1#2\hfil\cr}}}
\def\overleftarrow{\mathpalette\overleftarrow@}
\def\overleftarrow@#1#2{\vbox{\ialign{$##$\cr
     #1{\leftarrow}\mkern-6mu\cleaders\hbox{$#1\mkern-2mu{-}\mkern-2mu$}\hfill
      \mkern-6mu{-}\cr
     \noalign{\kern -1\p@\nointerlineskip}
     \hfil#1#2\hfil\cr}}}
\def\overleftrightarrow{\mathpalette\overleftrightarrow@}
\def\overleftrightarrow@#1#2{\vbox{\ialign{$##$\cr
     #1{\leftarrow}\mkern-6mu\cleaders\hbox{$#1\mkern-2mu{-}\mkern-2mu$}\hfill
       \mkern-6mu{\to}\cr
    \noalign{\kern -1\p@\nointerlineskip}
      \hfil#1#2\hfil\cr}}}
\def\underrightarrow{\mathpalette\underrightarrow@}
\def\underrightarrow@#1#2{\vtop{\ialign{$##$\cr
    \hfil#1#2\hfil\cr
     \noalign{\kern -1\p@\nointerlineskip}
    #1{-}\mkern-6mu\cleaders\hbox{$#1\mkern-2mu{-}\mkern-2mu$}\hfill
     \mkern-6mu{\to}\cr}}}
\def\underleftarrow{\mathpalette\underleftarrow@}
\def\underleftarrow@#1#2{\vtop{\ialign{$##$\cr
     \hfil#1#2\hfil\cr
     \noalign{\kern -1\p@\nointerlineskip}
     #1{\leftarrow}\mkern-6mu\cleaders\hbox{$#1\mkern-2mu{-}\mkern-2mu$}\hfill
      \mkern-6mu{-}\cr}}}
\def\underleftrightarrow{\mathpalette\underleftrightarrow@}
\def\underleftrightarrow@#1#2{\vtop{\ialign{$##$\cr
      \hfil#1#2\hfil\cr
    \noalign{\kern -1\p@\nointerlineskip}
     #1{\leftarrow}\mkern-6mu\cleaders\hbox{$#1\mkern-2mu{-}\mkern-2mu$}\hfill
       \mkern-6mu{\to}\cr}}}
\def\sqrt#1{\radical"270370 {#1}}
\def\dots{\relax\ifmmode\let\next=\ldots\else\let\next=\tdots@\fi\next}
\def\tdots@{\unskip\ \tdots@@}
\def\tdots@@{\futurelet\next\tdots@@@}
\def\tdots@@@{$\mathinner{\ldotp\ldotp\ldotp}\,
   \ifx\next,$\else
   \ifx\next.\,$\else
   \ifx\next;\,$\else
   \ifx\next:\,$\else
   \ifx\next?\,$\else
   \ifx\next!\,$\else
   $ \fi\fi\fi\fi\fi\fi}
\def\text{\relax\ifmmode\let\next=\text@\else\let\next=\text@@\fi\next}
\def\text@@#1{\hbox{#1}}
\def\text@#1{\mathchoice
 {\hbox{\everymath{\displaystyle}\def\textfonti{\the\textfont1 }%
    \def\textfontii{\the\textfont2 }\textdef@@ T#1}}
 {\hbox{\everymath{\textstyle}\def\textfonti{\the\textfont1 }%
    \def\textfontii{\the\textfont2 }\textdef@@ T#1}}
 {\hbox{\everymath{\scriptstyle}\def\textfonti{\the\scriptfont1 }%
   \def\textfontii{\the\scriptfont2 }\textdef@@ S\rm#1}}
 {\hbox{\everymath{\scriptscriptstyle}\def\textfonti{\the\scriptscriptfont1 }%
   \def\textfontii{\the\scriptscriptfont2 }\textdef@@ s\rm#1}}}
\def\textdef@@#1{\textdef@#1\rm \textdef@#1\bf
   \textdef@#1\sl \textdef@#1\it}

\def\textdef@#1#2{\def\next{\csname\expandafter\eat@\string#2fam\endcsname}%
\if S#1\edef#2{\the\scriptfont\next\relax}%
 \else\if s#1\edef#2{\the\scriptscriptfont\next\relax}%
 \else\edef#2{\the\textfont\next\relax}\fi\fi}
\scriptfont\itfam=\tenit \scriptscriptfont\itfam=\tenit
\scriptfont\slfam=\tensl \scriptscriptfont\slfam=\tensl
\mathcode`\0="0030
\mathcode`\1="0031
\mathcode`\2="0032
\mathcode`\3="0033
\mathcode`\4="0034
\mathcode`\5="0035
\mathcode`\6="0036
\mathcode`\7="0037
\mathcode`\8="0038
\mathcode`\9="0039
\def\Cal{\relax\ifmmode\let\next=\Cal@\else
     \def\next{\errmessage{Use \string\Cal\space only in math mode}}\fi\next}
\def\Cal@#1{{\fam2 #1}}
\def\bold{\relax\ifmmode\let\next=\bold@\else
   \def\next{\errmessage{Use \string\bold\space only in math
      mode}}\fi\next}\def\bold@#1{{\fam\bffam #1}}
\mathchardef\Gamma="0000
\mathchardef\Delta="0001
\mathchardef\Theta="0002
\mathchardef\Lambda="0003
\mathchardef\Xi="0004
\mathchardef\Pi="0005
\mathchardef\Sigma="0006
\mathchardef\Upsilon="0007
\mathchardef\Phi="0008
\mathchardef\Psi="0009
\mathchardef\Omega="000A
\mathchardef\varGamma="0100
\mathchardef\varDelta="0101
\mathchardef\varTheta="0102
\mathchardef\varLambda="0103
\mathchardef\varXi="0104
\mathchardef\varPi="0105
\mathchardef\varSigma="0106
\mathchardef\varUpsilon="0107
\mathchardef\varPhi="0108
\mathchardef\varPsi="0109
\mathchardef\varOmega="010A
\def\wlog#1{\immediate\write-1{#1}}
\catcode`\@=12  
\def\=def{\; \mathop{=}_{\text{\rm def}} \;}
\def\rd{\partial}
\def\MB{{\text{MB}}}
\def\PB{{\text{PB}}}
\def\ahat{\hat{a}}
\def\phat{\hat{p}}
\def\qhat{\hat{q}}
\def\Ahat{\hat{A}}
\def\Uhat{\hat{U}}
\def\Vhat{\hat{V}}
\def\What{\hat{W}}
\def\uhat{\hat{u}}
\def\vhat{\hat{v}}
\def\what{\hat{w}}
\def\calU{{\cal U}}
\def\calV{{\cal V}}
\def\calUhat{\hat{\calU}}
\def\calVhat{\hat{\calV}}
\def\calL{{\cal L}}
\def\calM{{\cal M}}
\def\calG{{\cal G}}
\def\Win{W_{\text{in}}}
\def\Whatin{\What_{\text{in}}}
%
\hsize=15.5truecm
\vsize=23truecm
\sequentialequations
\doublespace
\TagsOnRight
\overfullrule=0pt
\pubnum={Kyoto University KUCP-0054/92}
\date={December 1992}
\titlepage
\title{DRESSING OPERATOR APPROACH TO MOYAL ALGEBRAIC DEFORMATION
OF SELFDUAL GRAVITY}
\author{Kanehisa Takasaki}
\address{
  Faculty of Integrated Human Studies, Kyoto University\break
  Yoshida-Nihonmatsu-cho, Sakyo-ku, Kyoto 606, Japan\break
  E-mail: takasaki@ jpnyitp (Bitnet)\break
}
\abstract
\noindent
Recently Strachan introduced a Moyal algebraic deformation of
selfdual gravity, replacing a Poisson bracket of the Plebanski
equation by a Moyal bracket. The dressing operator method in
soliton theory can be extended to this Moyal algebraic
deformation of selfdual gravity. Dressing operators are defined
as Laurent series with coefficients in the Moyal (or star product)
algebra, and turn out to satisfy a factorization
relation similar to the case of the KP and Toda hierarchies.
It is a loop algebra of the Moyal algebra (i.e., of a $W_\infty$
algebra) and an associated loop group that characterize this
factorization relation. The nonlinear problem is linearized on
this loop group and turns out to be integrable.

\endpage

The notion of Moyal algebras
[\REF\Moyal{
    Moyal, J., Proc. Camb. Phil. Soc. 45 (1949), 99; \hfill\break
    Baker, G., Phys. Rev. 109 (1958), 2198; \hfill\break
    Fairlie,D.G., Proc. Camb. Phil. Soc. 60 (1964), 581; \hfill\break
    Bayen, F., Flato, M., Fronstal, C., Lichnerowicz, A., and
   Sternheimer, D., Ann. Phys. (N.Y.) 111 (1978), 61, 111; \hfill\break
    Averson, W., Commun. Math. Phys. Phys. 89 (1983), 77; \hfill\break
    Fletcher, P., Phys. Lett. B248 (1990), 323; \hfill\break
    Fairlie, D.B., Fletcher, P., and Zachos, C.K.,
    J. Math. Phys. 31 (1990), 1085.}\Moyal]
is a kind of quantum deformation of Poisson algebras. Since the
end of the eighties, there has been renewed interest in these
algebras. This is because in two dimensions, they give an explicit
realization of the two types of W-infinity algebras --- quantum
($W_\infty$) and classical ($w_\infty$) algebras. It is now widely
recognized that W-infinity algebras of both types are deeply
linked with integrability of nonlinear systems.

A number of integrable systems are now known to be related to
$w_\infty$ algebras. Even within the context of field theory,
one can pick out several important examples such as:
the dispersionless KP hierarchy
    [\REF\KoGidKP{Kodama, Y., Phys. Lett. 129A (1988), 223;\hfill\break
    Kodama, Y., and Gibbons, J., Phys. Lett. 135A (1989), 167.}
    \KoGidKP]%
    [\REF\Kr{Krichever, I.M., Commun. Math. Phys. 143 (1991), 415.}\Kr]%
    [\REF\TaTadKP{Takasaki, K., and Takebe, T.,
    Int. J. Mod. Phys. A7, Suppl. 1B (1992), 889.}\TaTadKP],
the SU($\infty$) Toda field theory
    [\REF\SaVe{Saveliev, M.V., and Vershik, A.M.,
    Commun. Math. Phys. 126 (1989), 367;
    Phys. Lett. 143A (1990), 121.}\SaVe]%
    [\REF\Ba{Bakas, I., Commun. Math. Phys. 134 (1990), 487.}\Ba]%
    [\REF\PaToda{Park, Q-H., Phys. Lett. 236B (1990), 429.}\PaToda],
its hierarchy (the dispersionless Toda hierarchy)
    [\REF\KoGidT{Kodama, Y., Phys. Lett. 147A (1990), 477.}\KoGidT]%
    [\REF\TaTadT{Takasaki, K., and Takebe, T.,
    Lett. Math. Phys. 23 (1991), 205.}\TaTadT],
the selfdual vacuum Einstein equation (selfdual gravity)
    [\REF\BoPl{Boyer, C.P., and Plebanski, J.F.,
    J. Math. Phys. 26 (1985), 229.}\BoPl]%
    [\REF\Ta{Takasaki, K., J. Math. Phys. 31 (1990), 1877.}\Ta]%
    [\REF\PaSDG{Park, Q-H., Phys. Lett. 238B (1990), 287;
    Int. J. Mod. Phys. A7 (1992), 1415.}\PaSDG]%
    [\REF\YaCh{Yamagishi, K., and Chapline, F.,
    Class. Quantum Grav. 8 (1991), 427.}\YaCh], etc.
In these integrable systems, $w_\infty$ algebras are realized as
the Poisson algebra Poisson($\Sigma$) or the algebra sdiff($\Sigma$) of
area-preserving diffeomorphisms on a two dimensional surface $\Sigma$.
Penrose's twistor theory
    [\REF\Pe{Penrose, R., Gen. Rel. Grav. 7 (1976), 31.}\Pe]
provides a unified framework for understanding integrability of
these systems.

One may naturally ask if these integrable systems of $w_\infty$ type
have any integrable deformation associated with a $W_\infty$ algebra.
The dispersionless KP and Toda hierarchies do have such a $W_\infty$
analogue, i.e., the ordinary KP and Toda hierarchies, which are
of course integrable. A $W_\infty$ analogue of selfdual gravity is
recently proposed by Strachan
    [\REF\St{Strachan, I.A.B., Phys. Lett. B282 (1992), 63.}\St]
as a Moyal algebraic deformation of the selfdual vacuum Einstein
equation. The problem of proving its integrability, however,
remains obscure.

In this paper, we consider this integrability problem by means of
soliton theoretical techniques rather than of twistor theory.
In soliton theory, the notion of ``dressing operators" plays a
central role. Our strategy is to construct dressing operators for
Strachan's deformation of selfdual gravity.  We will then derive
a ``factorization relation" that connects those dressing operators
with their ``initial values" on a two dimensional subspace of
space-time. This technique is borrowed from a similar approach to
the KP hierarchy
    [\REF\Mu{Mulase, M., Advances in Math. 54 (1984), 57.}\Mu]
and the Toda hierarchy
    [\REF\TaTaToda{Takasaki, K., in: {\it Group Representations and
    Systems of Differential Equations\/},
    Advanced Studies in Pure Math. 4
    (Kinokuniya, Tokyo, 1984); \hfill\break
    Takebe, T., Lett. Math. Phys. 21 (1991), 77.}\TaTaToda],
and as in those cases, we can thereby show that the nonlinear problem
is converted into a linear problem on an infinite dimensional group.

Let us recall that the Plebanski equation
    [\REF\Pl{Plebanski, J.F., J. Math. Phys. 16 (1975), 2395.}\Pl]
$$
  \{ \Omega_{,p}, \Omega_{,q}\}_\PB  \equiv
  \Omega_{,p\phat} \Omega_{,q\qhat}
  - \Omega_{,p\qhat} \Omega_{,q\phat}  = 1,    \tag\eqname\Plebanski
$$
where $\Omega$ is an unknown function (K\"ahler potential)
of suitable space-time coordinates $(p,q,\phat,\qhat)$,
gives a local expression of selfdual vacuum Einstein spaces.
Strachan's idea [\St] is to replace the Poisson bracket
$$
  \{ F, G \}_\PB
  = \frac{\rd F}{\rd \phat} \frac{\rd G}{\rd \qhat}
   -\frac{\rd F}{\rd \qhat} \frac{\rd G}{\rd \phat}  \tag\eq
$$
by the Moyal bracket
$$
  \{ F, G \}_\MB
  = \frac{2}{\hbar}\sinh \left[ \frac{\hbar}{2} \left(
        \frac{\rd^2}{\rd\phat \rd\qhat'}
       -\frac{\rd^2}{\rd\qhat \rd\phat'} \right) \right]
    \left. F(\phat,\qhat) G(\phat',\qhat')
           \right|_{\phat'=\phat,\qhat'=\phat}.      \tag\eq
$$
[This definition is somewhat unusual, but reproduces the
ordinary Moyal bracket if one replaces $\hbar \to i\hbar$.
This is harmless, because $\hbar$ in this note is just
a formal parameter. This substitution rule $\hbar \to i\hbar$
applies to all formulas in the following.] The deformed equation
$$
  \{ \Omega_{,p}, \Omega_{,q} \}_\MB = 1   \tag\eqname\MoyalPlebanski
$$
reduces to Eq. (\Plebanski) in the quasi-classical ($\hbar \to 0$)
limit.

Integrability of selfdual gravity is related to the existence of
auxiliary variables, so called ``twistor functions"
    [\REF\Ne{Newman, E.T., Porter, J.R., and Tod, K.P.,
    Gen. Rel. Grav. 9 (1978), 1129.}\Ne].
They  are functions of both space-time coordinates and a
``spectral parameter" $\lambda$, and define a correspondence
between the space-time and the twistor space. For local analysis
of selfdual gravity, it is sufficient to consider four such
functions $\calU,\calV,\calUhat,\calVhat$ [\BoPl][\Ta]. They
have Laurent series expansion
$$
\align
  &\calU = \lambda p + \sum_{n=0}^\infty u_n \lambda^{-n}, \quad
   \calV = \lambda q + \sum_{n=0}^\infty v_n \lambda^{-n}, \\
  &\calUhat = \phat + \sum_{n=1}^\infty \uhat_n \lambda^n, \quad
   \calVhat = \qhat + \sum_{n=1}^\infty \vhat_n \lambda^n,
                                                  \tag\eq  \\
\endalign
$$
and satisfy the ``Lax equations"
$$
\align
  & \frac{\rd \calU}{\rd p}
    + \lambda \{\frac{\rd\Omega}{\rd p}, \calU \}_\PB = 0, \quad
    \frac{\rd \calU}{\rd q}
    + \lambda \{\frac{\rd\Omega}{\rd q}, \calU \}_\PB = 0,   \\
  & \ldots (\text{same equations with}\ \calU \
            \text{replaced by}\ \calV,\calUhat,\calVhat) \ldots
                                                  \tag\eq  \\
\endalign
$$
and the ``canonical Poisson relations''
$$
    \{ \calU, \calV \}_\PB
  = \{ \calUhat, \calVhat \}_\PB = 1.                   \tag\eq
$$
The Plebanski equation, (\Plebanski), gives the Frobenius
integrability condition of these equations. These equations can
further be converted into the 2-form equation
$$
  d\calU \wedge d\calV = d\calUhat \wedge d\calVhat, \tag\eq
$$
which is a clue for understanding various aspects of integrability
of selfdual gravity [\BoPl][\Ta].

It is easy to see how these auxiliary variables can be extended to
the deformed equation. Moyal algebraic counterparts of
$\calU,\calV, \calUhat,\calVhat$, say $U,V,\Uhat,\Vhat$, will be
given by Laurent series of the form
$$
\align
  U = \lambda p + \sum_{n=0}^\infty u_n \lambda^{-n},  \quad &
  V = \lambda q + \sum_{n=0}^\infty v_n \lambda^{-n},         \\
  \Uhat = \phat + \sum_{n=1}^\infty \uhat_n \lambda^n, \quad &
  \Vhat = \qhat + \sum_{n=1}^\infty \vhat_n \lambda^n
                                                      \tag\eq \\
\endalign
$$
with $\hbar$-dependent coefficients that have smooth $\hbar \to 0$
limit as
$$
  u_n(\hbar,p,q,\phat,\qhat) = u_n^{(0)}(p,q,\phat,\qhat)
                               + O(\hbar),
  \ \text{etc} \ \ldots.                      \tag\eqname\FormOfUn
$$
The Moyal algebraic version of the Lax equations will be given by
$$
\align
  & \frac{\rd U}{\rd p}
    + \lambda \{\frac{\rd\Omega}{\rd p}, U \}_\MB = 0, \quad
    \frac{\rd U}{\rd q}
    + \lambda \{\frac{\rd\Omega}{\rd q}, U \}_\MB = 0,     \\
  & \ldots (\text{same equations with}\ U \
            \text{replaced by}\ V,\Uhat,\Vhat) \ldots.
                                                  \tag\eq \\
\endalign
$$
The canonical Poisson relations will be replaced by the
``canonical commutation relations"
$$
  \{ U, V \}_\MB = \{ \Uhat, \Vhat \}_\MB = 1.    \tag\eq
$$
One can indeed show that the deformed Plebanski equation,
(\MoyalPlebanski), becomes the Frobenius integrability condition
of these equations. Obviously, these equations reduce to the
previous equations in the quasi-classical ($\hbar\to 0$) limit.

It will be instructive to compare the present situation with the
relation between the KP hierarchy and its quasi-classical limit
(the dispersionless KP hierarchy) [\KoGidKP][\Kr][\TaTadKP].
The Lax formalism of the KP hierarchy consists of a set of Lax
equations for a canonical conjugate pair of pseudo-differential
operators $L$ and $M$, $[L,M]=1$. In the quasi-classical limit,
$L$ and $M$ are replaced by a canonical conjugate pair of
functions $\calL$ and $\calM$, $\{ \calL, \calM \} = 1$,
on a two dimensional phase space with a Poisson bracket
$\{\ , \ \}$. They satisfy a set of quasi-classical Lax
equations in which the commutator of pseudo-differential
operators is replaced by the Poisson bracket. Thus the Lax
formalism of the Plebanski equation is almost parallel to
the dispersionless KP hierarchy, both being related to a
Poisson algebra.  We expect a similar correspondence between
the Lax formalism of the Moyal algebraic Plebanski equation
and the KP hierarchy.

To make this analogy more precise, let us recall that the
Moyal bracket can be expressed as a (normalized) commutator
$$
  \{ F, G \}_\MB = \frac{2}{\hbar}( F*G - G*F)    \tag\eq
$$
of the star product [\Moyal]
$$
  F*G = \exp \left[ \frac{\hbar}{2} \left(
          \frac{\rd^2}{\rd \phat \rd \qhat'}
          -\frac{\rd^2}{\rd \qhat \rd \phat'} \right) \right]
        \left. F(\phat,\qhat) G(\phat',\qhat')
        \right|_{\phat'=\phat,\qhat'=\qhat}.       \tag\eq
$$
The star product is associative and non-commutative, and in fact
coincides with the composition rule of (pseudo)differential
operators in ``Weyl ordering" (hence gives a realization of
$W_\infty$ algebra). Hoppe et al.
    [\REF\Ho{Hoppe, J., Olshanetsky, M., and Theisen, S.,
    Kahlsruhe preprint KA-THEP-10/91 (October, 1991).}\Ho]
uses this associative algebraic structure behind the Moyal bracket
to study a family of integrable systems related to Moyal algebras.
Although those integrable systems are different from ours, this is
very suggestive. In our case, basic building blocks of the theory
are Laurent series with coefficients in the Moyal algebra, which
form a loop algebra of the Moyal algebra. The star product can be
extended to this loop algebra as
$$
    (F\lambda^n) * (G\lambda^m) =  F*G \lambda^{n+m}    \tag\eq
$$
and defines an associative algebraic structure.  In the case of
the KP hierarchy, the same role is played by pseudo-differential
operators, which also form an associative and non-commutative algebra.

These observations indicate us what a dressing operator approach to
the Moyal algebraic Plebanski equation looks like. The dressing
operator of the KP hierarchy is a pseudo-differential operator
of the form $W = 1 + w_1(\rd/\rd x)^{-1} + \cdots$.  Dressing
operators of the Moyal algebraic Plebanski equation should be
Laurent series of $\lambda$ with coefficients in the Moyal algebra.

More specifically, we need two dressing operators, say $W$ and
$\What$ to express the two different types of Lax operators
$(U,V)$ and $(\Uhat,\Vhat)$ in a dressing form,
$$
\align
  U = W*(\phat + p\lambda)*W^{-1}, \quad &
  V = W*(\qhat + q\lambda)*W^{-1},                         \\
  \Uhat = \What*\phat*\What^{-1}, \quad &
  \Vhat = \What*\qhat*\What^{-1}.
                                                 \tag\eq   \\
\endalign
$$
This is rather similar to the Toda hierarchy, and one will therefore
infer that these dressing relations can be satisfied by Laurent series
of the form
$$
\align
  W = \sum_{n=0}^\infty w_n \lambda^{-n},          \quad &
  w_0  \not= 0,                                           \\
  \What = 1 + \sum_{n=1}^\infty \what_n \lambda^n, \quad &
  \what_1 = - \Omega /\hbar
                                                   \tag\eq \\
\endalign
$$
with coefficients depending on $\hbar$ and $(p,q,\phat,\qhat)$.
This is, however, not yet enough if we take into account the
requirement that the coefficients $u_n$, etc. of the Lax operators
behave smoothly in the limit of $\hbar \to 0$ as shown in (\FormOfUn).
This requirement is fulfilled if $W$ and $\What$ are written
$$
\align
  W = \exp_{*} \hbar^{-1} A(\hbar,p,q,\phat,\qhat,\lambda),  \quad &
  A = \sum_{n=0}^\infty a_n(\hbar,p,q,\phat,\qhat)\lambda^{-n},     \\
  \What = \exp_{*} \hbar^{-1}
              \Ahat(\hbar,p,q,\phat,\qhat,\lambda),          \quad &
  \Ahat = \sum_{n=1}^\infty
                       \ahat_n(\hbar,p,q,\phat,\qhat)\lambda^n,
                                              \tag\eqname\WisExp \\
\endalign
$$
with coefficients $a_n$ and $\ahat_n$ that have smooth limit
as $\hbar \to 0$. If we select these dressing operators
appropriately, the Lax equations can be converted into
the evolution equations
$$
\align
  \hbar \frac{\rd W}{\rd p}
  =& -\lambda\frac{\rd \Omega}{\rd p}*W + W*\qhat\lambda,
                                                           \\
  \hbar \frac{\rd W}{\rd q}
  =& -\lambda\frac{\rd \Omega}{\rd q}*W - W*\phat\lambda,
                                                           \\
  \hbar \frac{\rd \What}{\rd p}
  =& \left( -\lambda\frac{\rd\Omega}{\rd p}
            -\half q\lambda^2 \right) * \What,
                                                           \\
  \hbar \frac{\rd \What}{\rd q}
  =& \left( -\lambda\frac{\rd\Omega}{\rd q}
            +\half p\lambda^2 \right) * \What
                                     \tag\eqname\FlowOfW \\
\endalign
$$
of the dressing operators. This is quite parallel to the dressing
operator formalism of the KP and Toda hierarchies, apart from the
strange extra terms $q\lambda^2/2$ and $p\lambda^2/2$. These
extra terms originate in non-commutativity of the two flows
($p$-flow and $q$-flow) as we shall see later.

We can now apply the factorization technique for the KP and Toda
hierarchies [\Mu][\TaTaToda] to our problem. Consequently,
it turns out that the dressing operators and their ``initial values"
$ \Win = W|_{p=q=0}$ and $ \Whatin = \What|_{p=q=0}$ are connected
by the factorization relation
$$
    W^{-1}*\What = e(p,q) * \Win^{-1}*\Whatin.
                                           \tag\eqname\Factorize
$$
The boost operator $e(p,q)$ of time evolution is given by
$$
    e(p,q) = \exp_{*}[ \hbar^{-1}( - p\qhat\lambda
                       + q\phat\lambda ) ],
                                                \tag\eq
$$
where $\exp_{*}$ is the star exponential,
$$
    \exp_{*} F = 1 + \sum_{n=1}^\infty \frac{1}{n!}
                      F*\cdots*F \quad (n\text{-fold product}).
                                                \tag\eq
$$
This factorization relation may be understood as an integrated form
of Eq. (\FlowOfW). If, conversely, one can solve this relation for
a given initial data, $W$ and $\What$ automatically satisfy
Eq. (\FlowOfW) hence give rise to a solution of the Moyal algebraic
Plebanski equation.

Solvability of this factorization problem is ensured, at least in
a neighborhood of $(p,q)=(0,0)$, by the following reasoning.
The set of Laurent series of the form $\hbar^{-1}A + \hbar^{-1}\Ahat$,
where $A$ and $\Ahat$ are as in Eq. (\WisExp), is closed under the
star product commutator, therefore gives a Lie algebra $\calG$ with
a natural direct sum decomposition into two Lie subalgebras,
$$
    \calG = \calG_{\le 0} \oplus \calG_{\ge 1},  \quad
    \hbar^{-1}A     \in \calG_{\le 0}, \quad
    \hbar^{-1}\Ahat \in \calG_{\ge 1}.                 \tag\eq
$$
This induces a decomposition at the group level,
$$
    \exp \calG = \exp \calG_{\le 0} \cdot \exp \calG_{\ge 1},
                                                       \tag\eq
$$
at least in a neighborhood of the identity element. Thus,
given initial data $\Win \in \exp \calG_{\le 0}$ and
$\Whatin \in \exp \calG_{\ge 1}$, one can find two factors
$W \in \exp \calG_{\le 0}$ and $\What \in \exp \calG_{\ge 1}$
that satisfy the factorization relation.

The Lie algebra $\calG$ is a kind of loop algebra of the Moyal
algebra (hence of a $W_\infty$ algebra). This gives a quantum
deformation of a similar loop algebra of the Poisson algebra
(i.e., of a $w_\infty$ algebra) in the ordinary Plebanski
equation [\BoPl][\Ta]. As we have seen, it is this loop algebra
rather than the Moyal algebra itself that characterizes diverse
hidden structures of the Moyal algebraic Plebanski equation.

Eq. (\FlowOfW) can now be interpreted as integrable flows on the
direct product group $\exp\calG_{\le 0} \times \exp\calG_{\ge 1}$.
As Mulase described impressively in the case of the KP hierarchy [\Mu],
the factorization relation now works as a machinery that links
the nonlinear world (the left hand side) with the linear world
(the right hand side) in which the flows are ``linearized"
into the left action of $e(p,q)$.  Furthermore, the two flows
are non-commutative, because the two terms in the star exponential
of $e(p,q)$ do not commute with respect to the star product.
To be more specific, $e(p,q)$ can be factorized as
$$
\align
    \exp_{*}[\hbar^{-1}( - p\qhat\lambda + q\phat\lambda)]
    =& \exp[\hbar^{-1}( - p\qhat\lambda - \half pq\lambda^2 )]
           *\exp[\hbar^{-1}q\phat\lambda]                       \\
    =& \exp[\hbar^{-1}( q\phat\lambda + \half pq\lambda^2 )]
           *\exp[-\hbar^{-1}p\qhat\lambda],
                                                     \tag\eq    \\
\endalign
$$
and thereby satisfies an unusual differentiation formula:
$$
\align
  \hbar \frac{\rd e(p,q)}{\rd p}
  =& (-\qhat\lambda -\half q\lambda^2) * e(p,q),
                                                     \\
  \hbar \frac{\rd e(p,q)}{\rd q}
  =& ( \phat\lambda +\half p\lambda^2) * e(p,q).
                                             \tag\eq \\
\endalign
$$
This is the origin of the extra terms $p\lambda^2/2$ and
$q\lambda^2/2$ in Eq. (\FlowOfW).

We have thus seen that the dressing operator method in soliton
theory can be extended to the Moyal algebraic Plebanski
equation.  Dressing operators are defined as Laurent series
with coefficients in the Moyal (or star product) algebra,
and turn out to satisfy a factorization relation
similar to the case of the KP and Toda hierarchies.
It is a loop algebra of the Moyal algebra (i.e., of a
$W_\infty$ algebra) and an associated loop group that
characterize this factorization relation. The nonlinear
problem is linearized on this loop group and turns out to
be integrable.

We add a few comments.

1. A hierarchy of higher flows can be constructed as in the case of
the KP and Toda hierarchies. One set of such flows are parametrized
by two series of variables $p_n, q_n, n=1,2,\ldots$ ($p_1=p$, $q_1=q$)
and generated by the star exponential
$$
  e(p_1,p_2,\ldots,q_1,q_2,\ldots)
  = \exp_{*}[\hbar^{-1}( -\sum_{n=1}^\infty p_n\qhat\lambda^n
                         +\sum_{n=1}^\infty q_n\phat\lambda^n )]
                                                  \tag\eq
$$
instead of the previous $e(p,q)$. Another set of flows with
variables $\phat_n, \qhat_n, n=1,2,\ldots$, which resemble
the ``negative flows" of the Toda hierarchy, are generated
by inserting a similar star exponential (but replacing
$\lambda^n \to \lambda^{-n}$) TO THE RIGHT SIDE of
$\Win^{-1}*\Whatin$ in Eq. (\Factorize).  The results of this
paper can be extended straightforward to these higher flows.

2. The factorization relation can also be used to construct a large
set of symmetries ($W_\infty$ symmetries). These symmetries are
generated by the action of a star product loop group element
from the left or right side of $\Win^{-1}*\Whatin$. The
aforementioned hierarchy of higher flows is just a subset of
these symmetries. In the quasi-classical ($\hbar \to 0$) limit,
these symmetries reproduce $w_\infty$ symmetries of the Plebanski
equation [\BoPl][\Ta].

3. The Plebanski equation is interpretated as the equation
of motion of a physical state in $N=2$ string theory
    [\REF\Oo{Ooguri, H., and Vafa, C., Nucl. Phys. B361 (1991), 469.}\Oo].
Can we find a similar interpretation of the Moyal algebraic
deformation?

4. It is also quite straightforward to extend our results
to higher ($2k$, $k=1,2,\ldots$) dimensional Moyal algebras.
This leads to a Moyal algebraic deformation of hyper-K\"ahler
geometry. Hyper-K\"ahler manifolds are known to give target
spaces of supersymmetric sigma models
    [\REF\Hi{Hitchin, N.J., Kahlhede, A., Lindstr\"om, U.,
    and Ro\v cek, M., Commun. Math. Phys. 108 (1987), 535.}\Hi].
Can we find a similar application of the Moyal algebraic
deformation of hyper-K\"ahler geometry?

The author is grateful to Jens Hoppe, Stanislav Pakuliak,
Ian Strachan, Takashi Takebe and Hiroshi Yamada for a number
of valuable comments and discussions. This work is partly
supported by the Grant-in-Aid for Scientific Research,
the Ministry of Education, Science and Culture, Japan.

\refout
\bye